\documentclass[reprint,aps,nofootinbib,superscriptaddress]{revtex4-2}
\usepackage{graphicx,amsmath,amsfonts,amssymb,slashed,float}
\usepackage{array,multirow}
\usepackage[utf8]{inputenc}
\usepackage[usenames,dvipsnames]{xcolor} 
\usepackage{dcolumn}
\usepackage{bbold,wasysym}
\usepackage{bm}
\usepackage[normalem]{ulem}
\usepackage[breaklinks=true,colorlinks=true,linkcolor=blue,urlcolor=blue,citecolor=blue,hyperfootnotes=false]{hyperref}

\DeclareUnicodeCharacter{3B5}{$\epsilon$}
\DeclareUnicodeCharacter{2212}{\ensuremath{-}}
\DeclareUnicodeCharacter{039B}{\ensuremath{\Lambda}}

\newcommand{\Euclid}{\emph{Euclid}}
\newcommand{\threetwopt}{3$\times$2pt}
\newcommand{\Planck}{\emph{Planck}}
\newcommand{\Swyft}{\texttt{Swyft}}
\newcommand{\lcdm}{$\Lambda$CDM}
\newcommand{\wcdm}{$w_0 w_a$CDM}

\newcommand{\thet}{{\bm{\theta}}}

\newcommand{\xx}{{\bm{x}}}
\newcommand{\yy}{{\bm{y}}}
\newcommand{\ab}{{\bm{a}}}
\newcommand{\bb}{{\bm{b}}}

\newcommand{\appref}[1]{\hyperref[#1]{App.~\ref*{#1}}}

\begin{document}

\preprint{APS/123-QED}

\title{How to embed any likelihood into SBI: \\ Application to \Planck ~+ Stage IV galaxy surveys and Dynamical Dark Energy}

\author{Guillermo Franco Abellán}
\email{g.francoabellan@uva.nl}
\affiliation{GRAPPA Institute, Institute for Theoretical Physics Amsterdam, \\University of Amsterdam, Science Park 904, 1098 XH Amsterdam, The Netherlands}
\author{Noemi Anau Montel}
\affiliation{Max-Planck-Institut für Astrophysik, Karl-Schwarzschild-Str.\ 1, 85748 Garching, Germany}
\author{Oleg Savchenko}
\affiliation{GRAPPA Institute, Institute for Theoretical Physics Amsterdam, \\University of Amsterdam, Science Park 904, 1098 XH Amsterdam, The Netherlands}
\author{Christoph Weniger}
\affiliation{GRAPPA Institute, Institute for Theoretical Physics Amsterdam, \\University of Amsterdam, Science Park 904, 1098 XH Amsterdam, The Netherlands}

\date{\today}

\begin{abstract} 
Simulation-based inference (SBI) allows fast Bayesian inference for simulators encoding implicit likelihoods. However, some explicit likelihoods cannot be easily reformulated as simulators, hindering their integration into combined analyses within SBI frameworks. One key example in cosmology is given by the \Planck~CMB likelihoods. We present a simple method to construct an effective simulator for any explicit likelihood using samples from a previously converged Markov Chain Monte Carlo (MCMC) run. This effective simulator can subsequently be combined with any forward simulator. To illustrate this method, we combine the full \Planck~CMB likelihoods with a \threetwopt~simulator (cosmic shear, galaxy clustering and their cross-correlation) for a Stage IV survey like \Euclid, and test evolving dark energy parameterized by the $w_0w_a$ equation-of-state. Assuming the \wcdm~cosmology hinted by DESI BAO DR2 + \Planck~2018 + PantheonPlus SNIa datasets, we find that  future \threetwopt~data alone could detect evolving dark energy at $5\sigma$, while its combination with current CMB, BAO and SNIa datasets could raise the detection to almost $7\sigma$. Moreover, thanks to simulation reuse enabled by SBI, we show that our joint analysis is in excellent agreement with MCMC while requiring \emph{zero} Boltzmann solver calls. This result opens up the possibility of performing massive global scans combining explicit and implicit likelihoods in a highly efficient way. 

\end{abstract}

\maketitle

\section{\label{sec:introduction}Introduction } 

The most stringent and robust cosmological results of the next decade will be obtained from the combination of multiple astronomical surveys. This includes data from past and current experiments like \Planck~\cite{Planck:2018vyg}, ACT \cite{ACT:2025fju}, SPT \cite{SPT-3G:2025bzu}, KiDS \cite{Wright:2025xka}, DES \cite{DES:2021wwk}, BOSS \cite{eBOSS:2020yzd}, and DESI \cite{DESI:2025zgx}, as well as forthcoming surveys such as \Euclid~\cite{Euclid:2024yrr}, LSST \cite{LSST:2008ijt}, Simons Observatory \cite{SimonsObservatory:2018koc}, and CMB-S4 \cite{CMB-S4:2016ple}. Performing Bayesian inference from these surveys typically involves evaluating a large number of theoretical models--each requiring costly calls to Boltzmann solvers like \texttt{CAMB} \cite{Lewis:1999bs} or \texttt{CLASS} \cite{Lesgourgues:2011re}--and marginalizing over numerous nuisance parameters. In the context of  Markov Chain Monte Carlo (MCMC) methods, this time consuming exercise needs to be repeated for each new combination of experiments,\footnote{To name just a few concrete examples, the recent DESI BAO DR2 analysis \cite{DESI:2025zgx} explored 14 configurations of experiments to test the \wcdm~model; the `$H_0$ Olympics' paper \cite{Schoneberg:2021qvd} analyzed 12 data combinations across 8 different `finalist' models; and 30 combinations of experiments were considered in \cite{Brinckmann:2018owf} to derive forecast constraints on neutrino masses.} which becomes increasingly challenging given the diversity and complexity of datasets. 

In recent years, simulation-based inference (SBI) \cite{Cranmer:2019eaq} has emerged as a powerful cosmological tool for Bayesian analysis in complex settings (see e.g.~\cite{SimBIG:2023ywd,DES:2024xij,vonWietersheim-Kramsta:2024cks, Novaes:2024dyh, Zeghal:2024kic, Nguyen:2024yth, Zubeldia:2025qlt, FrancoAbellan:2024tbj,Savchenko:2025jzs}).
By using forward simulations that encode implicit likelihoods, SBI offers several key benefits. First, it supports amortized inference for rapid re-analysis and cross-validation. Second, it can flexibly incorporate complex effects through forward modeling that would be expensive to compute or intractable to express analytically in a likelihood. Third, it can efficiently yield marginal parameter estimates by directly integrating out nuisance variables. Finally, because simulations can be reused across inference tasks, SBI eliminates the need for repeated model evaluations, greatly enhancing computational efficiency.

However, some legacy cosmological likelihoods, even if explicit in form, cannot be readily reformulated as forward simulators. A notable example are the \Planck~cosmic microwave background (CMB) likelihoods \cite{Planck:2019nip}. Specifically, the low-$\ell$ likelihoods (\texttt{Commander}, \texttt{SimAll}) are computed at the pixel map level due to the non-Gaussian nature of the power spectrum at large scales; and the high-$\ell$ likelihoods  (\texttt{Plik}) rely on ``pseudo-$C_\ell$'s'' from various frequency channels, which introduce 47 nuisance parameters to describe instrument noise and foregrounds. Such dataset is too complex to be simulated in large numbers, preventing its integration into SBI workflows and thus limiting the ability to leverage the benefits previously discussed. 

In this work, we present a simple yet general trick to transform any explicit likelihood into an \emph{effective simulator}. Our method introduces an \emph{auxiliary-observable}, derived from samples of a pre-converged MCMC run, to construct an effective simulator that faithfully reproduces the original explicit likelihood. This effective simulator can be directly used for SBI, inheriting the full suite of benefits offered by such framework. Specifically, from a single MCMC, one can rapidly generate many effective simulations, which can be reused across different SBI tasks. Moreover, SBI allows seamless combination of likelihoods, whether explicit ones (turned into effective simulators) or implicit. This directly addresses the computational challenge of performing massive global scans, as it enables inference without the need for additional model evaluations each time new data is added. 

Alternatively, our approach can  be used to speed up traditional sampling-based inference, by leveraging our effective simulator to train a fast likelihood emulator. This connects naturally to prior work on emulation in cosmology, where surrogate models are trained to approximate computationally expensive theory codes \cite{SpurioMancini:2021ppk,Nygaard:2022wri,Gunther:2025xrq} or likelihoods \cite{McClintock:2019ijs, Bevins:2022qsc} and then used as fast drop-in replacements in MCMC samplers (see also \cite{Mootoovaloo:2024sao,Taylor:2024eqc} for different methods to accelerate joint cosmological analyses). 

We validate our method through a series of cosmological applications. First, we build an effective \Planck~\lcdm~simulator and apply it within SBI, recovering posteriors that are in excellent agreement with the original MCMC. Next, we combine this effective simulator with a forecast simulator of \threetwopt~probes (weak lensing, galaxy clustering, and their cross-correlation) for a Stage IV photometric galaxy survey, and perform a joint \lcdm~analysis. Thanks to simulation reuse enabled by SBI, we find accurate posteriors for this data combination without requiring any new model evaluations. 

Finally, we demonstrate our approach on a directly relevant physics application, by extending the SBI analysis to probe dynamical dark energy parameterized by the Chevallier-Polarski-Linder (CPL) equation-of-state \cite{Chevallier:2000qy,Linder:2002et}. This is motivated by the recent exciting results by the DESI collaboration, which hint at departures from a cosmological constant \citep{DESI:2024mwx}. Specifically, the combination of data from CMB + DESI baryonic acoustic oscillation (BAO) + Type Ia supernovae (SNIa) shows a preference for CPL evolving dark energy over \lcdm~at the $\sim 3-4\sigma$ level, depending on the SNIa dataset used \cite{DESI:2025zgx}. This has sparked a strong debate in the community about the robustness of the results and potential implications for beyond-\lcdm~models \cite{Lewis:2024cqj,Nesseris:2025lke,Efstathiou:2025tie, Giare:2024oil, Santos:2025wiv, DESI:2025fii,Ye:2024ywg,Li:2024qus,Elbers:2024sha,Wolf:2025jlc,Nakagawa:2025ejs,Urena-Lopez:2025rad,Chen:2025mlf,Camarena:2025upt}. 

We explore the impact of  Stage IV photometric surveys like \Euclid~in light of the latest DESI results, as these surveys are expected to achieve percent-level constraints on the dark energy equation-of-state \cite{Euclid:2019clj}. To this end, we adopt the following strategy: we generate synthetic Stage IV photometric data assuming the CPL cosmology favored by present CMB+DESI BAO+SNIa data, and assess the resulting parameter uncertainties given different data configurations. Notably, we find that Stage IV photometric data alone could yield a $\sim 5\sigma$ detection of evolving dark energy, while its combination with current datasets could raise this significance to almost $7\sigma$. This underscores the crucial role that joint analyses will play in advancing our understanding of dark energy.

This paper is organized as follows. In \autoref{sec:methodology}, we present the methodology, describing the auxiliary-observable trick and the framework for combining explicit and implicit likelihoods. In \autoref{sec:cosmo_application} we detail the application to cosmology, starting with a description of the models and datasets considered, and then discussing the inference setup for both MCMC and SBI. In \autoref{sec:results}, we show our main results for the \lcdm~and CPL dark energy models. We conclude in \autoref{sec:conclusions}. Complementary information can be found in various appendices. In \appref{app:alternative} we propose an alternative formulation of the auxiliary-observable trick. In \appref{app:likelihood_emulator} we describe an application of our trick to build likelihood emulators that can be used for fast joint cosmological analyses. In \appref{app:coverage_test} we present the results of an empirical coverage test for the SBI results. Finally, in \appref{app:preference_CPL} we discuss our measure of the preference for evolving dark energy.

\section{\label{sec:methodology}Methodology}
 
\subsection{The auxiliary-observable trick }
\label{sec:trick_theory}

Our goal is to embed a given likelihood function into a simulation-based framework. We begin with a likelihood function $L(\thet) \equiv p(\xx_o \mid \thet)$, defined for an observation $\xx_o$, that we can evaluate for different parameter values $\thet$. The objective is to construct an effective simulator for an auxiliary-observable, dubbed $\ab$, that allows the use of SBI methods while preserving the information from our original likelihood and enabling inference on $\xx_o$. The construction works as follows.

First, we define a probability distribution that is proportional to our likelihood function:
\begin{equation}
    p_L(\thet) \propto L(\thet).
\end{equation}
This distribution essentially captures the shape of the likelihood in parameter space. In practice, samples from $p_L(\thet)$ can be obtained from previous Monte Carlo chains generated using the original likelihood $L(\thet)$ with wide uninformative priors over the parameters $\thet$.

The crucial step is to define the probability distribution of the auxiliary-observable. We do so via
\begin{equation}
    p(\ab \mid \thet) \equiv  p_L(\thet - \ab) \;.
\end{equation}
An important consequence of this definition is that, when evaluating the distribution at $\ab = {\bf{0}}$, we recover our original likelihood up to a normalizing factor:
\begin{equation}
    p(\ab = {\bf{0}} \mid \thet ) \propto L(\thet) \;.
\end{equation}
This property is what enables our construction to work.

To generate samples from this auxiliary-observable distribution, we use a simple procedure. For any given parameter $\thet$, we sample $\ab \sim p(\ab \mid \thet)$ by first drawing a random parameter $\thet'$ from the distribution $p_L(\thet')$ and then setting:
\begin{equation}
    \ab = \thet - \thet' \quad \text{with} \quad  \thet' \sim p_L(\thet') \text{ and } \thet \sim p(\thet) \;,  
\label{eq:a_definition}
\end{equation}
where $p(\thet)$ is our parameter prior. This sampling procedure yields pairs $(\ab,\thet)$ that follow the joint distribution $p(\ab,\thet)= p(\ab\mid\thet)p(\thet)$.

With these samples in hand, we can train any SBI algorithm -- whether neural posterior estimation (NPE) \cite{papamakarios_npe, Jeffrey:2020itg}, neural likelihood estimation (NLE) \cite{papamakarios_nle, Alsing:2019xrx}, or neural ratio estimation (NRE) \cite{Cranmer:2015bka,nre_louppe}. Thanks to this construction, when the trained SBI estimators are evaluated at $\ab = {\bf{0}}$, we recover the desired inference for the original observation $\xx_o$. For instance, in the case of a NPE network $\hat{q}_\phi(\thet \mid \ab)$, one obtains
\begin{equation}
    \hat{q}_\phi(\thet \mid \ab = {\bf{0}}) \simeq p(\thet \mid \xx_o) \;.
\end{equation}

Importantly, for a given dataset and cosmological model, a single MCMC run is sufficient to generate as many effective simulations as needed -- whether for inference on the same dataset or in combination with others, as will be detailed in the next subsection. 

\begin{figure}
\centering
\includegraphics[width=\linewidth]{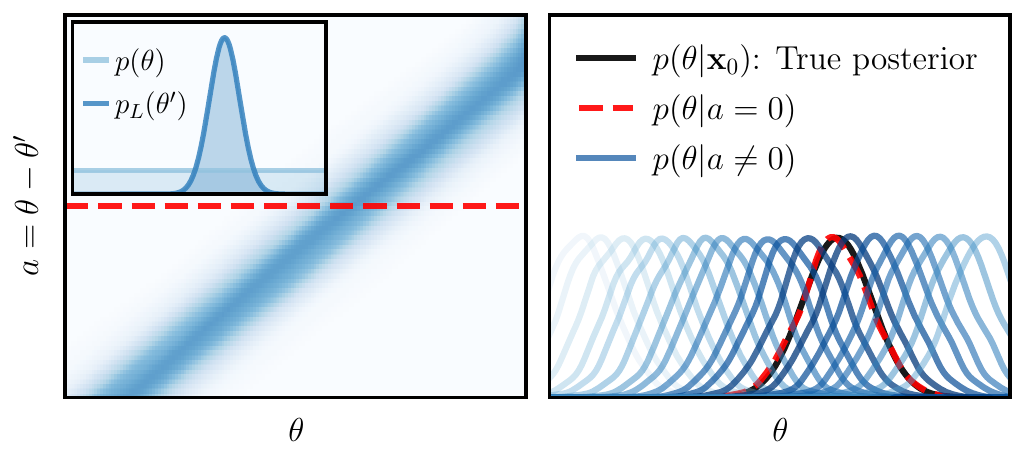}
\caption{Summary illustration of the auxiliary-observable construction for a toy 1-dimensional problem. Left panel: Definition of the auxiliary-observable $a$ according to \autoref{eq:a_definition} and its relationship with $\theta$. Right panel: Multiple conditional distributions $p(\theta\mid a)$ for different values of the auxiliary-observable $a$, with the true posterior $p(\theta\mid x_0)$ (black solid line) recovered for $a = 0$ (red dashed line). This shows how the framework preserves the original likelihood information at $a = 0$.}
\label{fig:toy}
\end{figure}

A visual summary of the auxiliary-observable trick is presented in \autoref{fig:toy} for a simple 1-dimensional case, illustrating how this construction preserves the original likelihood information when evaluated at $\ab={\bf{0}}$. An alternative formulation of the auxiliary-observable trick that provides additional intuition is presented in \appref{app:alternative}.

\subsection{Combining explicit and implicit likelihoods} \label{sec:combined}

The auxiliary-observable construction opens up the possibility of combining explicit and implicit likelihoods within a unified SBI framework. Consider a scenario where we have both a simulation model $p(\yy \mid \thet)$ (implicit likelihood) with observation $\yy_o$, and some additional likelihood constraint $L(\thet)$ (explicit likelihood) with observation $\xx_0$ that we want to combine into our analysis. This explicit likelihood can be converted into an effective simulator for $\ab$ following \autoref{eq:a_definition}, thereby enabling the use of simulations for both data sources.

We can then sample from the combined model:
\begin{equation} \label{eq:combined_sim}
    \ab, \yy, \thet \sim p(\ab \mid \thet) p(\yy \mid \thet) p(\thet) \;,
\end{equation}
where both $\ab$ and $\yy$ serve as training samples in our SBI runs. At inference time, to obtain joint constraints for observations $\xx_0$ and $\yy_0$, one just needs to evaluate the inference networks at $\ab = {\bf{0}}$ and $\yy = \yy_0$.

It is worth noting that the combined simulation model in \autoref{eq:combined_sim} assumes statistical independence between datasets, i.e. $p(\ab,\yy\mid\thet)=p(\ab\mid\thet)p(\yy\mid\thet)$.
This is usually a good approximation for many cosmological observables, and we adopt it throughout this work. As a real world application, in \autoref{sec:results} we will show examples combining the full \Planck~CMB likelihoods and a simulation model for a Stage IV photometric survey similar to \Euclid.\footnote{In principle, there exist correlations between CMB and large-scale structure (LSS) observables, arising mostly from secondary anisotropies of the CMB photons. We neglect these correlations as done in previous CMB+\Euclid~forecasts (e.g. \cite{Euclid:2024imf}), but we note that this correlation will be more important for a survey very sensitive to CMB lensing like CMB-S4 \cite{Kou:2025hvg}.} Given the independence assumption, this combined analysis could also be performed by simply generating simulations of Stage IV data with \Planck~priors. However, we emphasize that our method additionally allows to have \Planck~and Stage IV as separate simulators, that can be reused for multiple data combinations (this will be illustrated in \autoref{fig:num_sims}).

\section{\label{sec:cosmo_application}Application to cosmology}

\subsection{Models and data}

In this work, we analyze two different cosmological scenarios: the standard flat \lcdm~model, which assumes a cosmological constant dark energy with equation-of-state parameter $w = -1$; and the  \wcdm~model, which assumes a time-evolving dark energy with the CPL \cite{Chevallier:2000qy,Linder:2002et} parametrization for the equation-of-state parameter
\begin{equation}
w (a) = w_0 + w_a (1-a).
\end{equation}
To analyze the  \lcdm~model we use the parameter basis $\{H_0,100\omega_b,\omega_{\rm cdm},n_s, \ln{(10^{10}A_{\rm s})}, \tau_{\rm reio} \}$, where $H_0$ is the Hubble constant, $\omega_b$ and $\omega_{\rm cdm}$ are the physical baryon and cold dark matter abundances, $A_s$ and $n_s$ are the amplitude and tilt of the primordial power spectrum, and $\tau_{\rm reio}$ is the reionization optical depth. For \wcdm~we additionally vary the parameters $w_0$ and $w_a$, and use the parametrized post-Friedmann (PPF) approach \cite{Hu:2007pj,Fang:2008sn} to handle phantom crossing. To compute the quantities needed for the different cosmological observables (e.g., Hubble parameter $H(z)$, matter power spectrum $P_m(k,z)$, CMB anisotropy spectra $C_\ell^{XY}$) we use the public Boltzmann solver \texttt{CLASS}\footnote{\href{https://github.com/lesgourg/class_public}{\texttt{https://github.com/lesgourg/class\_public}}} \cite{Lesgourgues:2011re,Blas:2011rf} with the \texttt{halofit} prescription \cite{Smith:2002dz,Takahashi:2012em} for the non-linear corrections to the matter power spectrum. We adopt the \Planck~convention in modeling free-streaming neutrinos as two massless species and one massive with $m_\nu = 0.06 \ \rm{eV}$ \cite{Planck:2018vyg}. 

We will confront the \lcdm~and \wcdm~models against real data from CMB, BAO and SNIa, as well as synthetic data from an upcoming Stage IV photometric survey. We describe each of these in the following subsections.

\subsubsection{CMB, BAO and SNIa datasets}

For CMB, BAO and SNIa, we use the datasets detailed in the bullet points below. 

\begin{itemize}
    \item \textbf{\Planck~2018}: We consider the temperature (TT), polarization (EE) and cross (TE) power spectra from \Planck, specifically using the \texttt{Commander}, \texttt{SimAll} (for multipoles $\ell <30$) and \texttt{Plik}  (for multipoles $\ell \geq 30$) likelihoods, together with the lensing amplitude reconstruction from the official PR3 data release \cite{Planck:2019nip,Planck:2018lbu}. These likelihoods introduce 47 nuisance parameters to model instrument noise and foregrounds, of which we vary 21 following \texttt{Plik} recommendations.  
    
    \item \textbf{DESI BAO DR2}: We make use of the DESI BAO DR2 distance measurements from Table IV in \cite{DESI:2025zgx}: $D_V /r_d$ at $z = 0.295$; $D_M/r_d$ and $D_H/r_d$ (including their correlation) at $z = 0.51$, $0.706$, $0.934$, $1.321$, $1.484$ and $2.33$.

    \item \textbf{PantheonPlus}: We use the Pantheon+ catalog, which compiles information about the luminosity distance to over $1550$ SNIa in the redshift range $0.001 < z < 2.3$ \cite{Scolnic:2021amr, Brout:2022vxf}. This likelihood introduces one nuisance parameter $\mathcal{M}$ which describes the SNIa calibration.
\end{itemize}
In the following, we will use the data combination `\Planck~2018 + DESI~BAO~DR2 + PantheonPlus' (hereinafter referred to as `Baseline') to put constraints on \wcdm~model, whereas for \lcdm~model we will omit the DESI~BAO~DR2 and PantheonPlus datasets.  
This choice is motivated by the mild $\sim 2.3\sigma$ tension reported by DESI \cite{DESI:2025zgx} between their BAO data and the CMB within the \lcdm~model, a discrepancy which is alleviated for the \wcdm~model. 

\begin{table}
\begin{ruledtabular}
\begin{tabular}{cccccc} 
$z_m$ & $\bar{n}_g \ [\rm{arcmin}^{-2}]$  & $f_{\rm sky}$ & $\Delta \log_{10}\ell$ & $\sigma_\epsilon$ & $f_{\rm out}$ \\
 \colrule 
 $0.9$ & $30$  & $0.35$ & $0.08$ & $0.3$ & $0.1$  \\ 
 \colrule \noalign{\vskip 0.6mm} \colrule
 $c_b$ & $z_b$ & $\sigma_b$ & $c_o$ & $z_o$ & $\sigma_o$ \\ 
  \colrule 
  $1.0$ & $0$  & $0.05$ & $1.0$ & $0.1$ & $0.05$ \\ 
\end{tabular}
\end{ruledtabular}
\caption{Parameters describing the specifications for our Stage IV photometric survey. These include the median redshift of the survey $z_m$, the surface galaxy density $\bar{n}_g$, the sky fraction $f_{\rm sky}$ and the intrinsic ellipticity error $\sigma_\epsilon$. For definitions of the remaining parameters, we refer the reader to \cite{FrancoAbellan:2024tbj}.}
\label{tab:table_survey} 
\end{table}

\subsubsection{Synthetic \threetwopt~data}

On top of our \Planck/Baseline datasets, we consider data from a simulated \textbf{Stage IV photometric survey}. In particular, we focus on the so-called \threetwopt~signal, i.e.~the combination of weak lensing (WL), photometric galaxy clustering (GCph), and their cross-correlation. This is described by a series of angular power spectra $C_{ij}^{AB}(\ell)$, where $i$ and $j$ label different tomographic redshift bins, and $A$ and $B$ refer to either WL or GCph. For the construction of such observables from \texttt{CLASS} outputs, we use the same modeling approach as in \cite{FrancoAbellan:2024tbj}, and we refer the reader to that work for a detailed description. \ 

We consider $N_z = 10$ tomographic bins, and compute each spectrum $C_{ij}^{AB}(\ell)$ for $N_\ell = 29$ log-spaced values between $\ell_{\rm min}=10$ and  $\ell_{\rm max}=2000$.\footnote{We note that our choice $\ell^{\rm GC_{\rm ph}}_{\rm max}=\ell_{\rm max}^{\rm WL}=2000$ lies between the `pessimistic' ($\ell^{\rm GC_{\rm ph}}_{\rm max}=750, \ \ell_{\rm max}^{\rm WL}=1500$) and `optimistic' ($\ell^{\rm GC_{\rm ph}}_{\rm max}=3000, \ \ell_{\rm max}^{\rm WL}=5000$) configurations that have been usually considered in previous \Euclid~forecasts (e.g. \cite{Euclid:2020tff, Euclid:2021qvm, Euclid:2024imf}).} We assume these spectra to be distributed according to a multivariate Gaussian distribution; the parameters used to compute the data covariance matrix and the photometric redshift distributions are given in \autoref{tab:table_survey} (these correspond to the specifications for a Stage IV survey like \Euclid).  Our modeling of the \threetwopt~spectra necessitates the variation of 12 nuisance parameters: $A_{\rm IA}, \eta_{\rm IA}$, which describe intrinsic alignments, and $b_1,..., b_{10}$, which describe photometric galaxy bias (one per redshift bin).  \ 

To generate our synthetic dataset, we also need a fiducial cosmological model that we assume to be the one describing the Universe. Since we aim to combine mock \threetwopt~data with actual measurements from CMB, BAO and SNIa, this choice must be done with care---particularly because the maximum of the combined likelihoods may lie far from the fiducial point. Hence, we follow a similar strategy as the `fitted-Fisher approach' of \cite{Euclid:2021qvm}, and generate two sets of mock \threetwopt~data: one from the best-fit of \Planck~(used for the \lcdm~analysis) and another from the best-fit of our Baseline data (used for the \wcdm~analysis).  The corresponding fiducial values are reported in \autoref{tab:table_fid}. To obtain the best-fit parameters, we minimize the $\chi^2$ with the method explained in Appendix D.1 of \cite{Schoneberg:2021qvd}. For visualization purposes, we assume a `noiseless' \threetwopt~observation,  so that we obtain posteriors that are centered on the fiducial values.

\begin{table}
\begin{ruledtabular}
\begin{tabular}{ccc} 
\textrm{Parameter} & \textrm{Fiducial I} & \textrm{Fiducial II}\\
\colrule
$H_0$ [km s$^{-1}$ Mpc$^{-1}$] & 67.417  & 67.904  \\
$100~\omega_{\rm b}$           & 2.2392  & 2.2482  \\
$\omega_{\rm cdm}$             & 0.1199  & 0.1190  \\
$n_{\rm s}$                    & 0.9667  & 0.9691  \\
$\ln{(10^{10}A_{\rm s})}$      & 3.0452  & 3.0448  \\
$w_0$                          & -1.0    & -0.8276 \\
$w_a$                          & 0.0     & -0.6653 \\
\colrule
$A_{\rm IA}$                   & 1.72    & 1.72    \\
$\eta_{\rm IA}$                & -0.41   & -0.41   \\
$b_1$                          & 1.0998  &  1.0998 \\
$b_2$                          & 1.2203  &  1.2203 \\
$b_3$                          & 1.2724  &  1.2724 \\
$b_4$                          & 1.3166  &  1.3166 \\
$b_5$                          & 1.3581  &  1.3581 \\
$b_6$                          & 1.3998  &  1.3998 \\
$b_7$                          & 1.4447  &  1.4447 \\
$b_8$                          & 1.4965  &  1.4965 \\
$b_9$                          & 1.5653  &  1.5653 \\
$b_{10}$                       & 1.7430  &  1.7430 \\
\end{tabular}
\end{ruledtabular}
\caption{Fiducial cosmological and nuisance parameters values used to generate the mock \threetwopt~data. The cosmological parameters of `Fiducial I' correspond to the \lcdm~best-fit of \Planck~2018 data, while those  of `Fiducial II' correspond to the \wcdm~best-fit of our Baseline dataset  (\Planck~2018 + DESI BAO DR2 + PantheonPlus). The values for the nuisance parameters are the same as in \cite{FrancoAbellan:2024tbj}. }
\label{tab:table_fid}
\end{table}

\subsection{Inference setup}

We will confront each model against three different data combinations: 
\begin{itemize}
\item \lcdm: [\Planck, \threetwopt, \Planck+\threetwopt],
\item \wcdm: [Baseline, \threetwopt, Baseline+\threetwopt].
\end{itemize}
We perform each of these six different analyses using both MCMC and SBI (with the help of the auxiliary-observable trick). For the MCMC part, we use the public code \texttt{MontePython-v3}\footnote{\href{https://github.com/brinckmann/montepython_public}{ \texttt{https://github.com/brinckmann/montepython\_public}}} \cite{Audren:2012wb,Brinckmann:2018cvx}, where all the likelihoods described in the previous subsection have been implemented. We employ a Metropolis-Hastings (MH) algorithm assuming wide flat priors\footnote{We did not find the choice of prior boundaries to be particularly relevant for MCMC (as long as the posteriors never hit the prior), since the MH algorithm always explores points in the vicinity of the best-fit.} on the various cosmological parameters.  We deem chains to be converged with the Gelman-Rubin \cite{Gelman:1992zz} criterion $|R- 1| \leq 0.03$, and produce figures via the \texttt{GetDist}\footnote{\href{https://github.com/cmbant/getdist}{\tt https://github.com/cmbant/getdist}} package \cite{Lewis:2019xzd}. For the SBI analysis presented in the main body of the paper, we use an algorithm called  Marginal Neural Ratio Estimation (MNRE)~\cite{Miller:2021hys}, implemented via the open-source code \Swyft\footnote{\href{https://github.com/undark-lab/swyft}{\texttt{https://github.com/undark-lab/swyft}}}. MNRE uses simulated data-parameter pairs to train neural classifiers that directly learn 1-dimensional and 2-dimensional marginal posteriors of interest (see e.g. \cite{Cole:2021gwr, Montel:2022fhv,Karchev:2022xyn,Saxena:2023tue, Alvey:2023pkx,Bhardwaj:2023xph, Cole:2025sqo} for astrophysical or cosmological use cases). Like other SBI methods, MNRE requires two key components for each dataset: a forward simulator and a 
neural network design. The network is typically divided into two parts: the first performs a compression of the data into a small number of features, and the second carries out the actual inference based on the feature-parameter pairs. In the following, we describe these components for both the \threetwopt~and \Planck/Baseline~datasets. 

\begin{figure*}[ht!]
\includegraphics[scale=0.6]{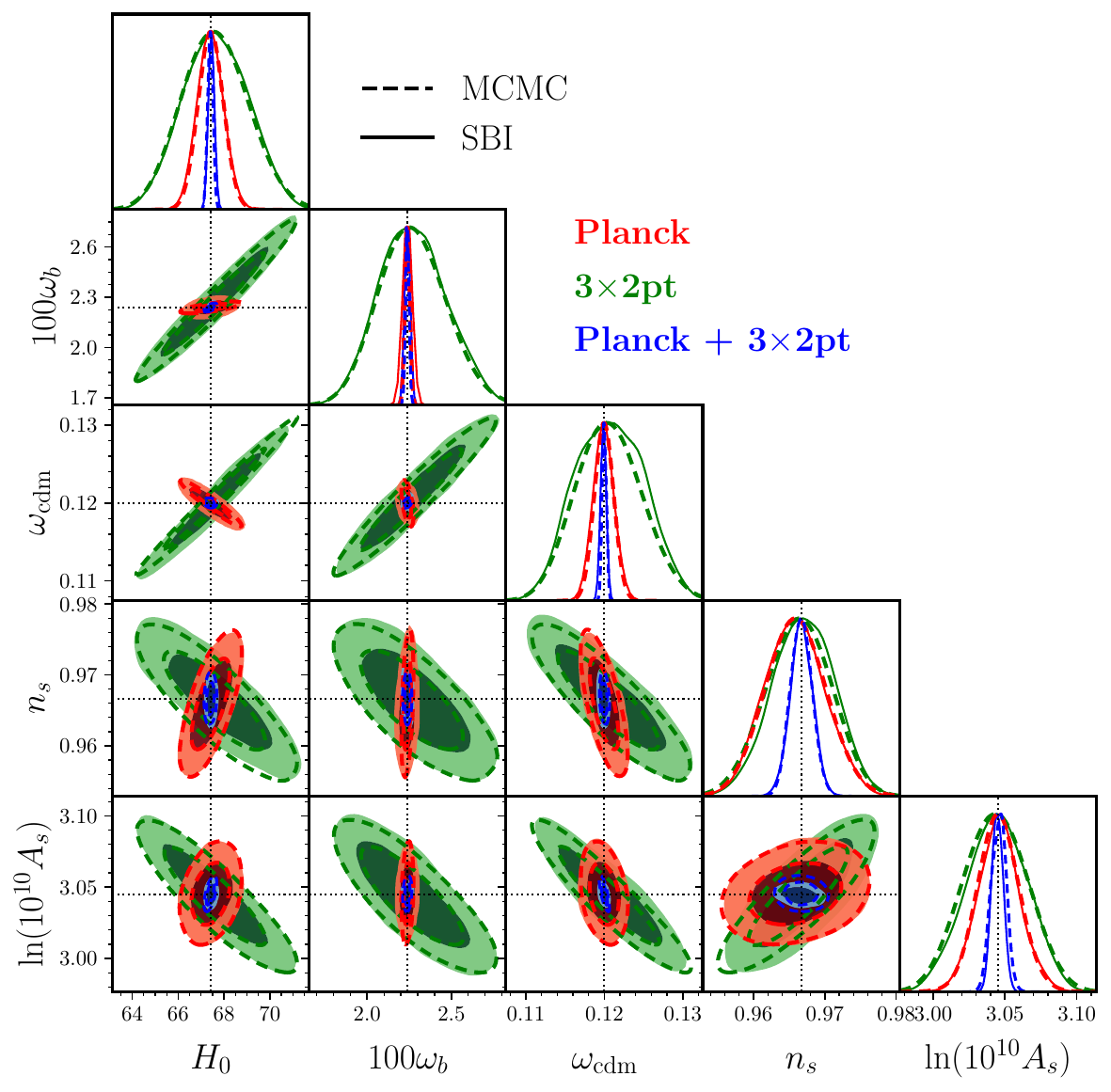}
\caption{1- and 2-dimensional marginalized posterior distributions (68\% and 95\% C.L.)  of the \lcdm~cosmological parameters for different data combinations, using both MCMC (dashed lines) and SBI (solid lines). The black dotted lines indicate the \lcdm~best-fit to \Planck~2018, which is our `Fiducial I' model used to generate the mock \threetwopt~observation for a Stage IV photometric survey. SBI posteriors are in excellent agreement with MCMC.}
\label{fig:lcdm_posteriors} 
\end{figure*}

\subsubsection{SBI pipeline for \threetwopt~data}

To perform inference from a Stage IV photometric survey, we employ the same forecast simulator of \threetwopt~power spectra that was used by \cite{FrancoAbellan:2024tbj}. We note that this simulator samples data vectors from a multivariate Gaussian with given covariance, making it equivalent to the explicit Gaussian likelihood that we incorporated in \texttt{MontePython-v3}. We do this for simplicity and to be able to cross-check our SBI results against MCMC, but we stress that the framework presented here would also allow to consider a more advanced simulator capturing any non-Gaussianities in the likelihood, like the one developed in \cite{vonWietersheim-Kramsta:2024cks}.\ 

We generate $5 \times 10^4$ simulations of \threetwopt~photometric spectra by varying the cosmological parameters  $\{H_0,100\omega_b,\omega_{\rm cdm},n_s, \ln{(10^{10}A_{\rm s})}\}$\footnote{It should be noted that \Euclid~probes are insensitive to $\tau_{\rm reio}$.} as well as the nuisance parameters $\{A_{\rm IA},\eta_{\rm IA},b_1,...,b_{10}\}$. When testing CPL dark energy, we generate another set of $5 \times 10^4$ simulations where we additionally vary $\{w_0,w_a\}$.\footnote{This number of simulations is not determined by a formal convergence criterion, such as the Gelman-Rubin diagnostic used for MCMC. Instead, it was empirically found to work well in previous MNRE analyses of \threetwopt~data \cite{FrancoAbellan:2024tbj}, a result we confirmed through comparison with MCMC (see \autoref{sec:results}).}
Each of these simulation batches was generated in parallel in  less than 2 hours using 72 CPU cores. We define the prior region based on the results of two preliminary Fisher analyses of \threetwopt~(one for \lcdm~and another for \wcdm), following a similar approach to that of \cite{FrancoAbellan:2024tbj}. In particular, each parameter $\theta_i$ is drawn from a uniform prior 
\begin{equation}
\theta_i \sim \mathcal{U}([\theta^0_i-3\sigma_i^F,\theta^0_i+3\sigma_i^F]),
\label{eq:prior_range}
\end{equation}
where $\theta^0_i$ denotes the fiducial values in \autoref{tab:table_fid} and the errors $\sigma_i^F$ are estimated from the corresponding Fisher matrix $F_{ij}$ as $\sigma_i^F=\sqrt{(F^{-1})_{ii}}$. For the parameters $w_0$ and $w_a$, we adopt slightly broader priors,
\begin{equation}
\theta_i \sim \mathcal{U}([\theta^0_i-5\sigma_i^F,\theta^0_i+5\sigma_i^F]).
\label{eq:prior_range_2}
\end{equation}
The priors in \autoref{eq:prior_range} might look relatively narrow from the point of view of a \threetwopt-only analysis, but these are usually much broader than \Planck-based posteriors (as we will see in \autoref{fig:lcdm_posteriors}). We also remark that very wide priors can easily be accommodated in SBI using sequential methods, which apply active learning techniques to quickly zoom into the relevant region of the parameter space \cite{papamakarios_nle,Cole:2021gwr}. 

For the data reduction of \threetwopt~power spectra, we employ the same strategy as in \cite{FrancoAbellan:2024tbj}, which combines Cholesky decomposition, principal component analysis (PCA), and linear compression to produce parameter-specific data summaries. These features are concatenated with the corresponding parameters, and fed as input to the ratio estimators, in charge of estimating every 1- and 2-dimensional marginal posterior. 
Throughout this work, each ratio estimator is implemented as a multi-layer perceptron (MLP) consisting of four residual blocks with 128 neurons each. 

\begin{figure*}[ht!]
\includegraphics[scale=0.45]{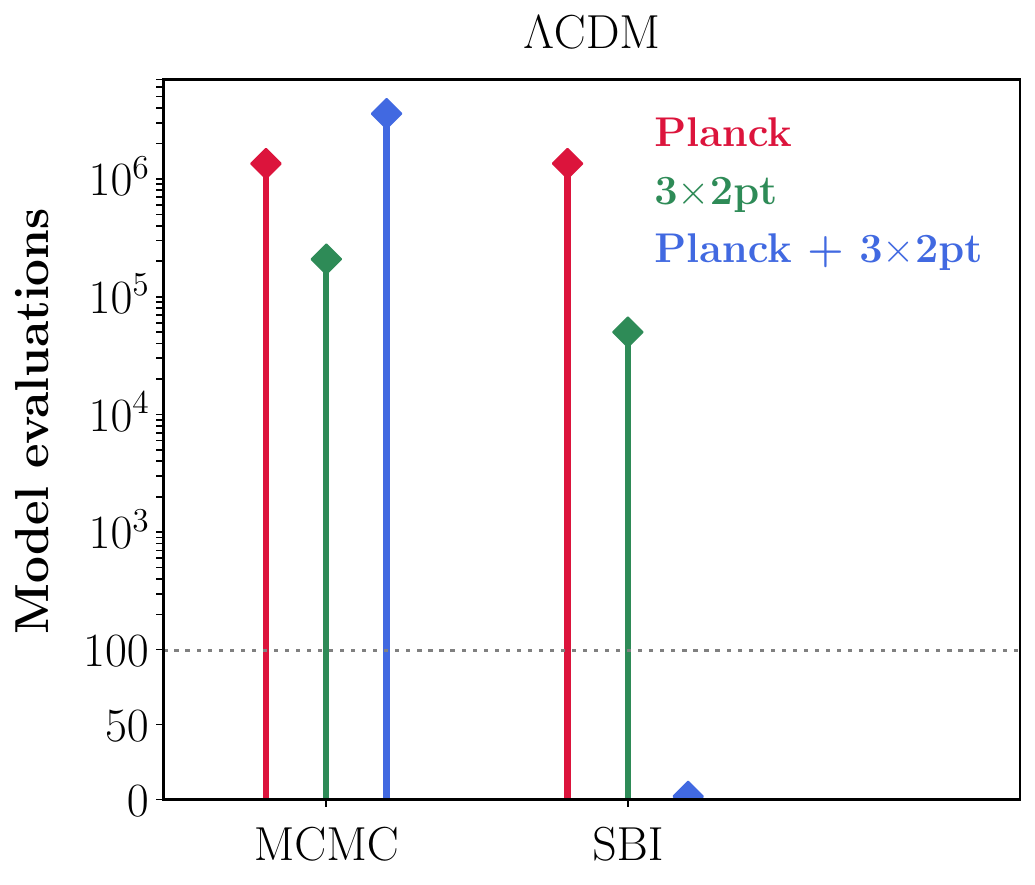}
\hspace{5mm}
\includegraphics[scale=0.45]{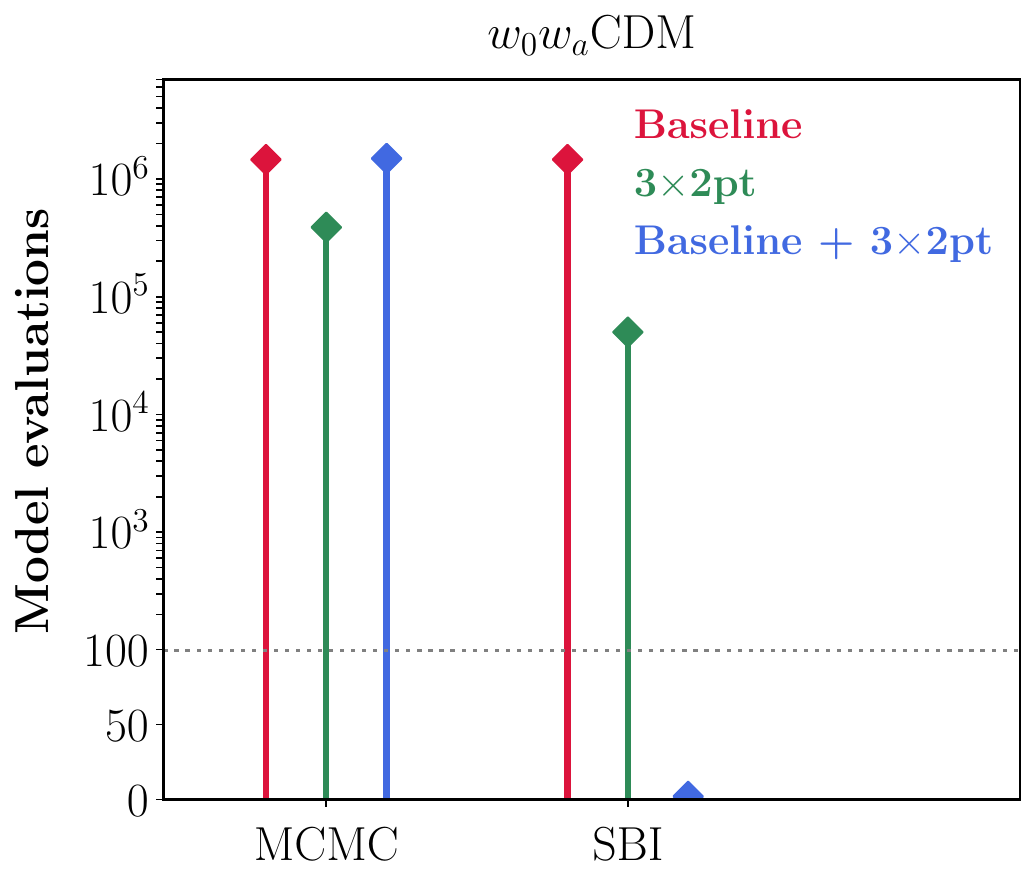}
\caption{Number of required model evaluations across all cosmological scenarios, datasets  and inference methods considered in this work. For MCMC, this reflects the number of likelihood evaluations needed for convergence, whereas for SBI, the meaning is different for each simulator. On the one hand, our effective simulators of \Planck~2018 and Baseline (= \Planck~2018 + DESI BAO DR2 + PantheonPlus) datasets use $10^5$ posterior samples that can be promptly generated from the corresponding pre-converged MCMC runs, and hence necessitate the same number of evaluations as MCMC. On the other hand, our forward \threetwopt~simulators use only $5\times 10^4$ data realizations, a factor $\sim 6$ smaller than those needed for MCMC. Finally, for the SBI combined analyses, we reused the samples from the \Planck/Baseline and \threetwopt~runs, hence requiring \emph{zero} new model evaluations. All SBI runs additionally involve training the inference networks, which takes just $5-20$min on a single GPU. 
}
\label{fig:num_sims}
\end{figure*}

\subsubsection{SBI pipeline for Planck/Baseline datasets}

For the \Planck~and Baseline datasets, we construct effective simulators using the trick described in \autoref{sec:trick_theory}. Namely,  we generate samples of the auxiliary-observable $\ab$ 
simply by drawing parameters $\thet$ from the prior and substracting random samples $\thet'$ from the corresponding MCMC run. To get each sample $\thet'$, we read the accepted points and weights stored in the MCMC files (excluding the burn-in phase), and then randomly select a point using the weights as probabilities. In this way, we build an effective \lcdm~simulator for \Planck, and an effective \wcdm~simulator for the Baseline combination.\footnote{The auxiliary-observable trick is technically not necessary for the DESI BAO DR2 and PantheonPlus datasets, as these are described by simple Gaussian likelihoods, which a priori could be easily reformulated as simulators. However, since we decided to analyze these likelihoods always in combination with \Planck, we found easier to apply the trick to the full Baseline dataset.} 

We generate $10^5$ samples from our effective \Planck~and Baseline simulators, by varying the cosmological parameters with the same priors as in \autoref{eq:prior_range}-\autoref{eq:prior_range_2}. We emphasize that these samples can be generated almost instantaneously, since they just rely on samples from a pre-converged MCMC. \ 

The auxiliary samples for \Planck/Baseline have already the same dimensionality as the parameters of interest, so the pre-compression step is not required. However, we still found it useful to standardize the samples using online normalization \cite{chiley2019onlinenormalizationtrainingneural}.
When incorporating the effective \Planck-based simulators in the combined analyses, we simply concatenate the auxiliary samples with the features extracted from the \threetwopt~spectra within each ratio estimator.\

\subsubsection{Training strategy}

For all the SBI runs carried out in this work, we allocate $80\%$ of the simulations for training, and the remaining $20\%$ for validation.  We use the Adam optimizer with a batch size of $256$ and an initial learning rate of $10^{-3}$. The learning rate is reduced by a factor of 0.1 if the validation loss plateaus for 3 consecutive epochs, and training is stopped if no improvement is observed for 5 epochs. Training for each inference run takes $\sim 5-20 \rm{\ min}$ on a single 40 GB \texttt{Nvidia A100} GPU.

\section{\label{sec:results}Results}

\subsection{\label{sec:results_lcdm}\lcdm~analysis}

In \autoref{fig:lcdm_posteriors} we show the cosmological constraints from the \lcdm~analysis of \Planck~2018,  \threetwopt~Stage IV photometric data, and their combination. We compare the results obtained with MCMC (dashed) against those obtained with SBI (solid). The dotted lines indicate the \lcdm~best-fit to \Planck, used to generate the mock \threetwopt~observation. In \autoref{fig:num_sims} we additionally show the number of model evaluations needed for all the analyses carried out in this work. \ 

We find that the SBI and MCMC posteriors are always in excellent agreement, with both methods yielding unbiased parameter constraints. Importantly, the agreement observed at the level of \Planck~contours (red) confirms the effectiveness of the auxiliary-observable trick that we introduced in \autoref{sec:methodology}, that is, our effective \Planck~simulator preserves the original likelihood information. 

A natural caveat is that constructing our effective simulator requires having first the corresponding MCMC runs. However, it is crucial to emphasize that this step needs to be performed only once per theoretical model, after which the resulting simulator can be reused for multiple data combinations. In practice, this permits a massive reduction in the number of simulations for combined analyses. Namely, for our SBI combined analysis (blue dashed), we could recycle both the $5\times 10^4$ simulations of \threetwopt~spectra (already a factor $\sim 6$ smaller than those needed for MCMC) as well as the \Planck~effective samples, hence allowing us to perform inference with \emph{zero} additional model evaluations. On the contrary, the MCMC combined analysis (blue solid) necessitated roughly $\sim 3\times 10^6$ likelihood evaluations to achieve convergence.\footnote{Among all the MCMC runs we performed, the combined \Planck+\threetwopt~\lcdm~analysis was notably slower to converge due to a poor prior knowledge of the covariance matrix used for the initial proposal distribution.}\ 

We stress that SBI is not always guaranteed to be substantially more simulator-efficient than MCMC for a given dataset. Indeed, the number of required simulations to achieve neural network convergence strongly depends on the chosen SBI algorithm and the network architecture, and a small simulation budget may lead to wrong posterior estimates \cite{Homer:2024cwg,Bairagi:2025sux}.
The key takeaway from \autoref{fig:num_sims} is rather that \emph{combined} analyses can greatly benefit from SBI, since simulations can be recycled across different dataset combinations, a feature unavailable in MCMC. Moreover, doing inference within SBI brings additional advantages, like the possibility to perform  statistical consistency checks which are usually unfeasible for MCMC. We illustrate this point in \appref{app:coverage_test}, where we conduct an empirical coverage test to further check the statistical consistency of the trained network.\ 

Even if our trick was originally designed to integrate arbitrary likelihoods into SBI, we demonstrate in \appref{app:likelihood_emulator} that it can alternatively be used to construct likelihood emulators for accelerating joint MCMC analyses.

\subsection{\wcdm~analysis}

We now turn to the case of CPL evolving dark energy. As we mentioned in \autoref{sec:introduction}, our goal is to find the significance with which Stage IV photometric surveys such as \Euclid~could detect the \wcdm~cosmology favored by the combination of DESI BAO, CMB and SNIa data. We also want to show how such kind of joint analysis, traditionally carried out with sampling-based methods, can be done efficiently with SBI, thanks to our auxiliary-observable trick. \ 

\begin{figure}[t!]
\includegraphics[scale=0.5]{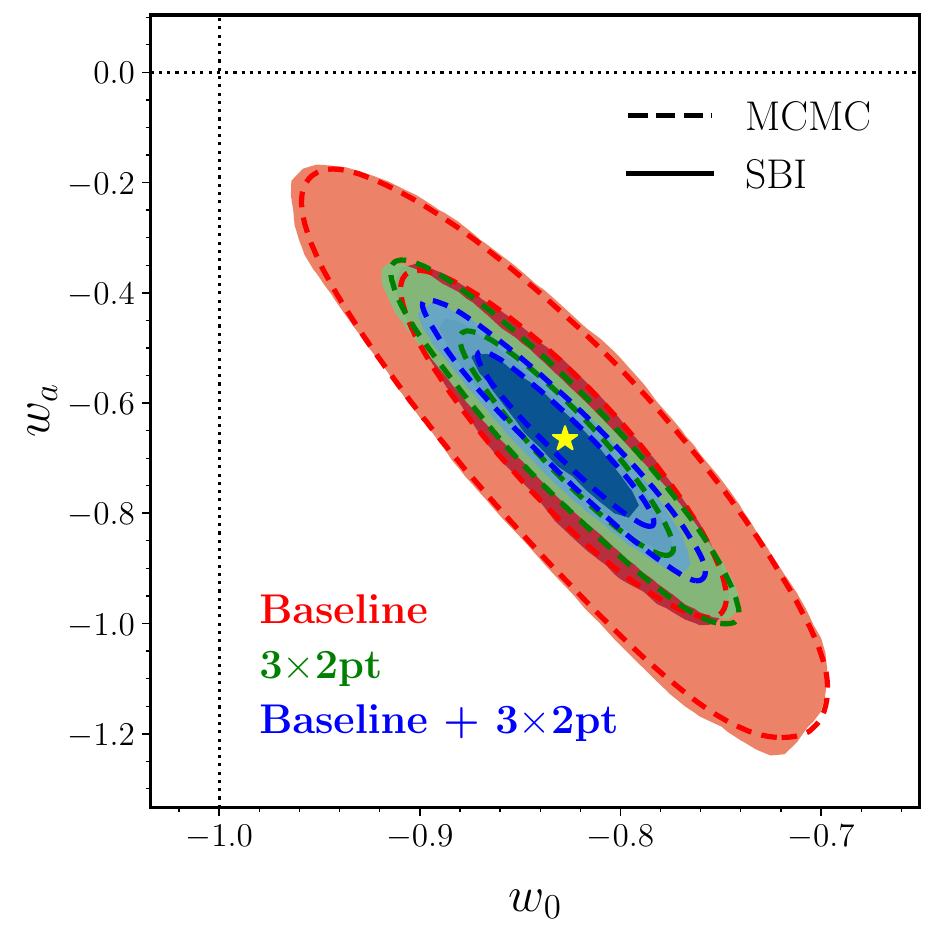} 
\caption{2-dimensional marginalized posterior (68\% and 95\% C.L.) of $w_a$ and $w_0$ for different data combinations, using both MCMC (dashed lines) and SBI (solid lines). The star indicates the \wcdm~best-fit to our Baseline dataset  (\Planck~2018 + DESI BAO DR2 + PantheonPlus), which is our `Fiducial II' model used to generate the mock Stage IV \threetwopt~observation. The black dotted lines indicate $w_0 = −1$ and $w_a = 0$; the \lcdm~limit lies at their intersection. The significance of rejection of \lcdm~is $3.3\sigma$, $5.0\sigma$ and $6.8\sigma$ for Baseline, \threetwopt~and their combination, respectively. 
}
\label{fig:w0wa_posteriors}
\end{figure}

In \autoref{fig:w0wa_posteriors}, we show the cosmological constraints on the $w_0$-$w_a$ plane for the different data combinations and the different methods (MCMC, SBI) considered in this paper. The star indicates the \wcdm~best-fit to our Baseline data, used to create the synthetic \threetwopt~observation. We observe that SBI posteriors are again in excellent agreement with MCMC. As for \lcdm, we could exploit simulation reuse to perform the Baseline+\threetwopt~combined analysis without needing any additional model evaluations (this is illustrated in the right panel of \autoref{fig:num_sims}).\ 

\begin{table}[t!]
\begin{ruledtabular}
\begin{tabular}{ccc} 
\textrm{Data set} & \textrm{MCMC} & \textrm{SBI}\\
\colrule
Baseline         & $3.31\sigma$  & $3.29\sigma$  \\
\threetwopt          &  $5.0\sigma$  & $4.99\sigma$ \\
Baseline+\threetwopt & $6.80\sigma$  & $6.82\sigma$ \\
\end{tabular}
\end{ruledtabular}
\caption{Significance of the detection of \wcdm~relative to \lcdm, estimated as the Mahalanobis distance $Q_\thet$ between our fiducial point $(w_0, w_a) = (-0.8276, -0.6653)$ and  $(w_0, w_a) = (-1, 0)$, for the various data sets and methods considered in this work.}
\label{tab:table_significances}
\end{table}

Regarding the scientific conclusion, our Baseline dataset indicates a preference for the \wcdm~ model over \lcdm~at the $\sim 3.3\sigma$ level (see \appref{app:preference_CPL} for details about our measure of the statistical significance of preference for evolving dark energy). This is comparable to the $\sim 2.8\sigma$ preference reported by the DESI collaboration using similar datasets\footnote{We note that the baseline analysis by the DESI collaboration relies on newer CMB data. In particular, instead of using \texttt{Plik}, they adopt \texttt{Camspec} \cite{Rosenberg:2022sdy}, which is built on the \texttt{NPIPE} PR4 data release from \Planck, along with CMB lensing data from ACT DR6 \cite{ACT:2023kun}. However, the constraints on the $w_0$-$w_a$ plane were found very consistent using \texttt{Camspec} and \texttt{Plik} \cite{DESI:2025zgx}.} \cite{DESI:2025zgx}. Interestingly, we find that \threetwopt~data from an \Euclid-like survey alone could detect the best-fit \wcdm~model at the $\sim 5\sigma$ level, while the full combination of all data could raise the detection to the $\sim 6.8\sigma$ level. We remark that these numbers may slightly vary depending on the choice of $\ell_{\rm max}$ or the non-linear prescription applied to model weak lensing cosmic shear \cite{Euclid:2020tff}. However, we expect that our main conclusion, namely that Stage IV photometric measurements can detect dynamical dark energy hinted by DESI with high significance, to remain valid under small variations in the \threetwopt~modeling. We also emphasize that it is the full combination of cosmic shear and photometric galaxy clustering spectra that provides sufficient constraining power to robustly measure dynamical dark energy. Indeed, it was recently shown in \cite{SpurioMancini:2024qic} that Stage IV cosmic shear alone could distinguish evolving dark energy preferred by DESI only at the field-level, but not at the power spectrum level.

\section{\label{sec:conclusions}Conclusions and outlook}

In this paper, we have introduced a simple yet effective way to incorporate arbitrary explicit likelihoods into modern simulation-based analysis frameworks. This is particularly relevant for legacy cosmological likelihoods such as \Planck, which cannot be easily reformulated as physical forward simulators. The approach, presented in \autoref{sec:methodology}, is based on constructing effective simulators from pre-converged MCMC runs, enabling the combination of explicit and implicit likelihoods within an unified SBI analysis pipeline. Alternatively, these effective simulators can be used to accelerate joint constraints in classical sampling-based inference by training fast likelihood emulators, as shown in \appref{app:likelihood_emulator}.\

Importantly, for a given theoretical model, a single converged MCMC is needed to define such an effective simulator, which can subsequently be reused across different data combinations. Moreover, converged chains are routinely released by major survey collaborations alongside their data products (see e.g. \cite{Addison:2019rqn}), providing a ready resource for constructing effective simulators without requiring access to the original likelihood code. This makes it possible to perform global fits very efficiently, bypassing expensive model evaluations for every new experimental configuration.

We validated our method by conducting several cosmological applications, in which each analysis was performed using both MCMC and SBI, and the resulting posteriors were compared. First, we did the analysis for the standard \lcdm~model, using the full \Planck~2018 likelihoods, synthetic \threetwopt~measurements for a Stage IV photometric survey, and the combination of both datasets. 
Within SBI, the \Planck~and \threetwopt~probes were modeled as effective and forward simulators, respectively. 
We found excellent agreement between MCMC and SBI ({shown in \autoref{fig:lcdm_posteriors}), showcasing the effectiveness of our method for combining forward simulators with explicit likelihoods, even when a generative model of the latter is not available. Furthermore, we could exploit simulation reusability to perform the SBI joint analysis at \emph{zero} simulator cost, greatly reducing computational time.\

Next, we applied the previous analysis setup to the \wcdm~evolving dark energy model, and extended \Planck~with BAO data from DESI DR2 and SNIa data from PantheonPlus (constituting our baseline dataset). Again, we found very good overlap between SBI and MCMC results (shown in \autoref{fig:w0wa_posteriors}), thus demonstrating the applicability of our method to efficiently test \lcdm~extensions. Our results indicate that Stage IV \threetwopt~data alone could potentially raise the detection of evolving dark energy hinted by DESI from $\sim 3\sigma$ to $\sim 5\sigma$, and up to $\sim 7\sigma$ when combined with the baseline dataset.\ 

Several clear improvements could extend the reach of our proposed framework. For instance, the current approach assumes statistical independence between datasets, which is a relatively common approximation in cosmological global fits. However, correlations are known to exist between certain cosmological probes, which may affect joint constraints. For future experiments, a way forward would be that major collaborations release realistic joint simulators for their various observables, such that correlations are automatically modeled. For past observations, one could simply construct joint likelihoods that properly account for the cross-correlations, and then apply the auxiliary-observable trick to preserve this structure within SBI analyses. 

Another caveat is that our effective simulators require separate Monte Carlo chains for different theoretical models. This could quickly become a burden if one wants to explore a large number of cosmological scenarios, which has become a growing practice given the accumulation and persistence of observational discrepancies in cosmology (see e.g. \cite{CosmoVerse:2025txj} for a recent review). Hence, developing more model agnostic techniques could help reduce this computational overhead.

Nonetheless, the framework represents a significant advance, as it brings legacy data into alignment with the capabilities of modern SBI techniques, and it allows efficient combinations of multiple cosmological datasets.

\begin{acknowledgments}
We thank Fabian Schmidt for his helpful comments on the draft. GFA, OS and CW acknowledge support from the European Research Council (ERC) under the European Union's Horizon 2020 research and innovation programme (Grant agreement No. 864035 - Undark). The main analysis for this work was carried out on the Snellius Computing Cluster at SURFsara.
\end{acknowledgments}

\section*{Data availability}

Accompanying code is available at \url{https://github.com/GuillermoFrancoAbellan/SBI-aux-obs}.%
\hspace{1mm}The data underlying this article will
be shared on reasonable request to the corresponding author. 

\appendix

\section{Alternative formulation} \label{app:alternative}

We present here an alternative formulation of the auxiliary-observable trick that provides additional insight into the construction. Rather than working directly with the auxiliary-observable $\ab$ as in the main text, this formulation frames the problem in terms of a simple simulator model with additive noise.

Our goal remains to embed a given likelihood function $L(\thet)$ into a simulation-based framework. To this end, we define an auxiliary simulator model $p(\mathbf y \mid \thet)$, and observed data $\mathbf y_{\rm obs}$, such that $p(\mathbf y_{\rm obs} \mid \thet) \propto L(\thet)$.  For a given likelihood function, many different simulator models exist that fulfill this requirement. 

We focus here on the arguably simplest version of an auxiliary simulator model, which features homoscedastic (parameter-independent) noise $\mathbf n \sim p_N(\mathbf n)$, and a trivial dependence of $\mathbf y$ to the model parameter $\thet$.  That model takes the form
\begin{equation}
    \mathbf y = \thet + \mathbf n\;.
\end{equation}
In order to relate $\mathbf y_{\rm obs}$ and $p_N(\mathbf n)$ to the likelihood function, we define a probability distribution that is proportional to our likelihood function:
\begin{equation}
    p_L(\thet) \propto L(\thet)\;.
\end{equation}
If we then define the auxiliary-observable as the mean value of samples from $p_L(\thet)$,
\begin{equation}
    \mathbf y_{\rm obs} \equiv \mathbb E_{p_L(\thet)} [ \thet ]
\end{equation}
and the (homscedastic) noise as centered and inverted samples from $p_L$,
\begin{equation}
    p_N(\mathbf n) \equiv p_L(\mathbf y_{\rm obs} - \mathbf n)\;,
\end{equation}
it is straightforward to show, by using the definitions of the various distributions, that
\begin{equation}
    p(\mathbf y = \mathbf y_{\rm obs} \mid \thet) \propto L(\thet)\;.
\end{equation}
This is the desired result.

Given samples from $p_N(\mathbf n)$, we can generate samples from $p(\mathbf y \mid \thet)$ for arbitrary values of $\thet$.  Those noise samples, as well as the value of $\mathbf y_{\rm obs}$, can be generated using MCMC techniques.  We first generate MCMC samples for $\thet$ from $p_L(\thet)$.  Based on those, we (a) estimate $\mathbf y_{\rm obs}$ (the mean value of $\thet$) and (b) generate noise samples $\mathbf n \sim p_N(\mathbf n)$ by centering the MCMC samples around zero (subtracting $y_{\rm obs}$).

With these samples in hand, we can train any SBI algorithm, as described in the main text. Thanks to this construction, when we evaluate the trained SBI estimators at $\mathbf y = \mathbf y_{\rm obs}$, we recover the desired inference for our original observation $\xx_o$. For instance, in the case of a NPE network $\hat{q}_\phi(\thet \mid \mathbf y)$, one obtains
\begin{equation}
    \hat{q}_\phi(\thet \mid \mathbf y = \mathbf y_{\rm obs}) \simeq p(\thet \mid \xx = \xx_o) \;.
\end{equation}

\section{Using the effective simulations to build a likelihood emulator} \label{app:likelihood_emulator}

\begin{figure*}[ht!]
\includegraphics[scale=0.45]{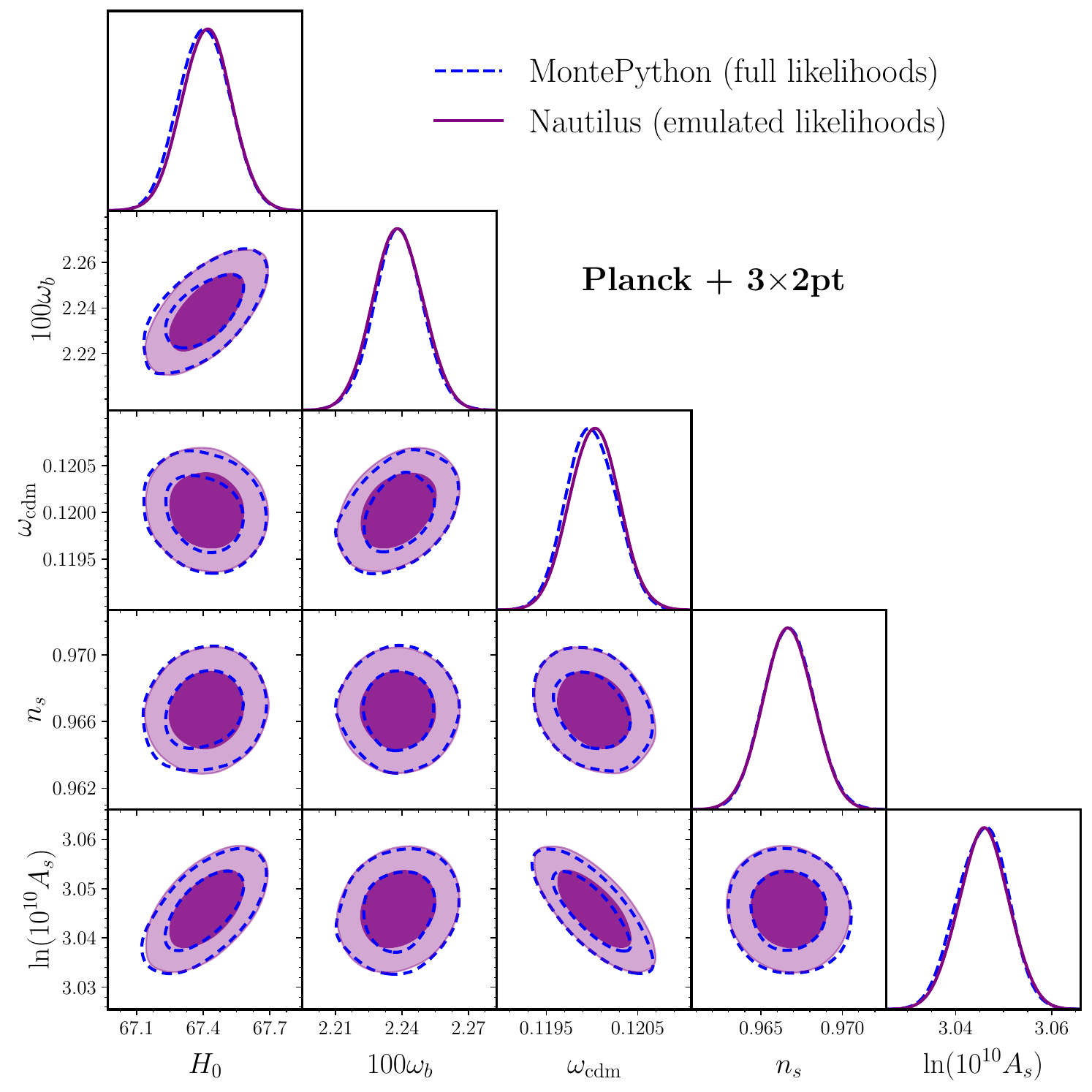}
\caption{ 1- and 2-dimensional marginalized posterior distributions (68\% and 95\% C.L.)  of the \lcdm~cosmological parameters for the combination of \Planck~and mock \threetwopt~data for a Stage IV photometric survey. These were obtained using both the MCMC sampler \texttt{MontePython-v3} with the full likelihoods (dashed blue lines) and the nested sampler \texttt{Nautilus} with the emulated likelihoods (solid purple lines). The posteriors from the emulated likelihoods are in excellent agreement with MCMC.}
\label{fig:likelihood_emulators}
\end{figure*}

As briefly mentioned in \autoref{sec:introduction}, a natural application of our effective simulator is to train a fast likelihood emulator for sampling-based techniques. This approach offers a complementary pathway to the full SBI framework described in the main text, allowing us to leverage the computational benefits of the auxiliary-observable method -- such as avoiding costly Boltzmann solver calls -- while working with traditional MCMC inference pipelines. In this appendix, we demonstrate this application through a combined \lcdm~analysis of \Planck~and Stage IV~\threetwopt~data, in a setup identical to the one described in \autoref{sec:cosmo_application}.\ 

Our approach involves training separate normalizing flows to emulate the likelihood for each dataset using NLE. For the \Planck~likelihood $p(\xx_0 \mid \thet)$, we train a normalizing flow $\hat{q}_\phi(\ab \mid \thet)$ on effective simulations $(\ab, \thet)$ constructed from a previous \Planck~MCMC run as described in \autoref{sec:trick_theory}. Similarly, for the Stage IV \threetwopt~likelihood $p(\yy_0 \mid \thet)$, we train a second normalizing flow $\hat{q}_\phi(\bb \mid \thet)$ on effective simulations $(\bb, \thet)$ constructed from a former \threetwopt~MCMC run. The mock \threetwopt~observation $\yy_0$ is the same that we used in \autoref{sec:results_lcdm}.\

We use the package \texttt{sbi}\footnote{ \href{https://github.com/sbi-dev/sbi}{ \texttt{https://github.com/sbi-dev/sbi}}} \cite{Tejero-Cantero2020} to train the normalizing flows. For both flows, we employ a Masked Autoregressive Flow \cite{Papamakarios:2017tec} architecture with 64 hidden features and 3 transformations. We use a learning rate of $10^{-3}$ and a batch size of $256$, training on $10^5$ effective simulations for each dataset. To standardize the input we use online normalization \cite{chiley2019onlinenormalizationtrainingneural}; without this normalization, the normalizing flows struggles to learn the likelihood surfaces accurately enough. \ 

Once trained, the likelihood emulators are obtained by evaluating the networks at the zero value of the auxiliary-observable. This evaluation recovers the original likelihood functions: $\hat{q}_\phi(\ab = {\bf{0}} \mid \thet) \simeq p(\xx_o \mid \thet)$ for \Planck~and $\hat{q}_\phi(\bb = {\bf{0}} \mid \thet) \simeq p(\yy_o \mid \thet)$ for the mock \threetwopt~data. The combined log-likelihood is then constructed as the sum of the two individual log-likelihood emulators, assuming independence between the CMB and LSS datasets as discussed in the main text (see \autoref{sec:combined}). 

For sampling this joint likelihood, we use the \texttt{Nautilus}\footnote{ \href{https://github.com/johannesulf/nautilus}{ \texttt{https://github.com/johannesulf/nautilus}}} nested sampler \cite{Lange:2023ydq}. Our results, shown in \autoref{fig:likelihood_emulators}, indicate that the posterior contours using the emulated likelihoods (solid purple lines) overlap almost perfectly with the original MCMC (dashed blue lines), while taking only a fraction of the total CPU time. This excellent agreement validates both the accuracy of our auxiliary-observable construction for different SBI algorithms, as well as the reliability of the trained emulators.  

Our likelihood emulation approach differs from most existing methods in cosmology, which typically focus on accelerating theoretical predictions like the matter power spectrum or CMB power spectra that feed into likelihood calculations
(e.g., \cite{SpurioMancini:2021ppk, Nygaard:2022wri,Gunther:2025xrq}).
In contrast, our method directly emulates `nuisance-marginalized' likelihoods from effective simulations, similar to the method proposed in \cite{Bevins:2022qsc}. Related techniques to accelerate joint likelihood-based analyses can be found in \cite{Mootoovaloo:2024sao,Taylor:2024eqc}.

\section{Coverage tests\label{app:coverage_test}}

\begin{figure*}[ht!]
\includegraphics[scale=0.45]{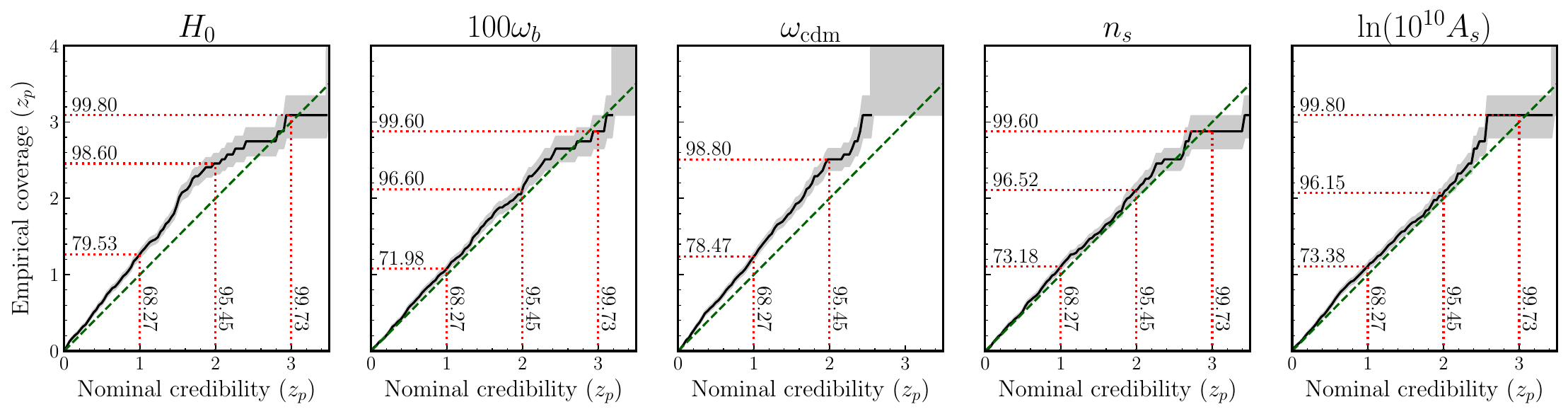 }
\caption{Coverage test for the cosmological parameters from the \lcdm~analysis of \Planck+3x2pt Stage IV photometric probes. The black solid line shows the average coverage, while the gray shaded region corresponds to the $68\%$ uncertainty on the coverage. The empirical coverage and confidence level generally match to good precision.}
\label{fig:coverage}
\end{figure*}

A key advantage of many SBI methods is that the trained inference networks are \emph{amortized}, i.e. they can estimate the posterior not only for the actual observation (as in MCMC), but also for \emph{any} mock observation simulated from the prior. This allows to asses fundamental statistical properties such as the \emph{empirical coverage} \cite{Hermans:2021rqv}, which gives the proportion of time that a given interval contains the true parameter value.  For a well-calibrated posterior, the $p\%$ credible interval should contain the simulation-truth value $p\%$ of the time, so one should get a perfectly diagonal line when plotting the empirical coverage against the expected credible levels. Crucially, our auxiliary-observable framework enables such coverage tests because it provides an effective simulator that allows training of amortized posteriors. Once trained on effective simulations, these posteriors can be evaluated at any mock auxiliary-observable. Such tests are generally unfeasible for standard likelihood-based methods, where one would need to run hundreds of separate MCMC runs -- one per each mock data simulation used in the coverage plot.

We perform a coverage test for the marginal \lcdm~posteriors of the \Planck+\threetwopt~analysis, using a batch of $500$ simulations. Instead of showing the highest posterior density region $p$, we use a different variable $z_p$ defined by $p/100 = \frac{1}{\sqrt{2\pi}}\int_{-z_p}^{z_p}dz \exp{(-z^2/2)}$, to put more emphasis in the tail of the posteriors. Hence, the usual $(1, 2, 3)\sigma$ regions correspond to $z_p = (1, 2, 3)$ with $p = (68.27, 95.45, 99.97)$. In addition, we compute the uncertainty on the empirical coverage arising from the finite number of samples, using the Jeffreys interval \cite{Cole:2021gwr}. The results of the coverage test are shown in \autoref{fig:coverage}. We find that the empirical coverage and the confidence levels generally match to good precision.

\section{Statistical significance of  \\ preference for evolving dark energy} \label{app:preference_CPL}

We aim to quantify the preference of CPL dark energy across the various data combinations considered in this work. Model selection criteria such as the Akaike Information Criterion (AIC) \cite{Akaike:1974vps} or the Bayesian Information Criterion (BIC) \cite{Schwarz:1978tpv} allow to quantify the improvement in the fit and penalize model complexity, and are computationally less demanding than Bayesian approaches like the Bayes factor ratio \cite{Kass:1995loi}. However, these criteria usually rely on differences in $\chi^2$ at the maximum a posteriori (MAP) points, which may be misleading when applied to synthetic data sets such as our mock \threetwopt~observation. Hence, we instead estimate the preference as the Mahalanobis distance in the 2-dimensional $w_0$-$w_a$ parameter space, which serves as a meaningful metric given that our posteriors are nearly Gaussian. In particular, we compute
\begin{equation}
Q_\thet = \sqrt{ (\vec{w}_{\rm Fid} -\vec{w}_{\Lambda})^T \bm{\Sigma}^{-1}_{w_0w_a} (\vec{w}_{\rm Fid} -\vec{w}_{\Lambda})  },
\label{eq:Q_theta}
\end{equation}
where $\vec{w}_{\rm Fid} = (w_0, w_a) = (-0.8276, -0.6653)$ is our fiducial point,  $\vec{w}_{\Lambda} = (w_0, w_a) = (-1, 0)$ is the \lcdm~limit,  and $\bm{\Sigma}_{w_0w_a} $ is the parameter covariance matrix reconstructed from the analysis. The metric $Q_\thet$ in \autoref{eq:Q_theta} generalizes the ``rule of
thumb difference in mean'' \cite{Raveri:2018wln} to various parameters, while incorporating the correlations between them. In \autoref{tab:table_significances} we report the values of $Q_\thet$ for the different data sets (Baseline, \threetwopt, Baseline+\threetwopt) and methods (MCMC, SBI) employed in the \wcdm~analysis. We find excellent agreement between the $Q_\thet$ values obtained using MCMC and SBI, which was expected given the good overlap of their corresponding contours shown in \autoref{sec:results}. 

\bibliography{references}

\end{document}